\documentclass[Print]{isecure-v24}

\usepackage{lipsum}
\usepackage{algpseudocode,algorithm}
\usepackage[colorlinks, bookmarksnumbered, linkcolor=darkblue, citecolor=darkblue, urlcolor=darkgreen]{hyperref}
\usepackage[numbers,sort&compress]{natbib}
\usepackage[all]{hypcap}
\usepackage{picins}
\usepackage{tikz}
\usetikzlibrary{shapes.geometric, arrows.meta, positioning, fit, backgrounds}

\usepackage{lettrine}
\usepackage{wrapfig}
\usepackage{graphicx}
\usepackage{subcaption}

\hypersetup{ colorlinks, citecolor=green, linkcolor=red, urlcolor=blue}
\usepackage{amsthm}
\usepackage{amssymb,amsmath,amsfonts,eqnarray,breqn}
\usepackage{fancyhdr}
\usepackage{ragged2e}
\usepackage{enumitem}   
\usepackage{refcount}
\usepackage[protrusion=true,expansion=true]{microtype}
\usepackage{float}
\usepackage[greek,english]{babel}
\usepackage[LGR,T1]{fontenc}

\usepackage{listings}
\usepackage{xcolor}
\usepackage{tikz}
\usetikzlibrary{arrows.meta, shapes, positioning, calc}

\usepackage{bookmark}

\DisableLigatures[f]{encoding=*,family=*}
\graphicspath{{Images/},{}}
\DeclareGraphicsExtensions{.jpg,.pdf,.png}
\input{Theorem-Styles.tex}
\articletype{Short Paper}

\usepackage{titlesec}
\titleformat{\paragraph}
{\normalfont\normalsize\bfseries}{\theparagraph}{1em}{}
\titlespacing*{\paragraph}
{0pt}{3.25ex plus 1ex minus .2ex}{1.5ex plus .2ex}

\journal{ISeCure}
\company{ISC}
\copyrightyear{2025}
\journalsite{http://www.isecure-journal.org}
\firstpage{1} 

\makeatletter
\begingroup \lccode`+=32 \lowercase
 {\endgroup \def\Url@ObeySp{\Url@Edit\Url@String{ }{+}}}
 \def\Url@space{\penalty\Url@sppen\ }
\makeatother

\usepackage[displaymath]{lineno}  
\newcommand*\patchAmsMathEnvironmentForLineno[1]{
  \expandafter\let\csname old#1\expandafter\endcsname\csname #1\endcsname
  \expandafter\let\csname oldend#1\expandafter\endcsname\csname end#1\endcsname
   \renewenvironment{#1}
     {\linenomath\csname old#1\endcsname}
     {\csname oldend#1\endcsname\endlinenomath}}
\newcommand*\patchBothAmsMathEnvironmentsForLineno[1]{
  \patchAmsMathEnvironmentForLineno{#1}
  \patchAmsMathEnvironmentForLineno{#1*}}
\AtBeginDocument{
\patchBothAmsMathEnvironmentsForLineno{equation}
\patchBothAmsMathEnvironmentsForLineno{align}
\patchBothAmsMathEnvironmentsForLineno{flalign}
\patchBothAmsMathEnvironmentsForLineno{alignat}
\patchBothAmsMathEnvironmentsForLineno{gather}
\patchBothAmsMathEnvironmentsForLineno{multline}
}

\begin{document}
\begin{frontmatter}


\def\NoDingTitle{Learning to Locate: GNN-Powered Vulnerability Path Discovery in Open Source Code}

\title{\NoDingTitle\textsuperscript{ }}


\author{Nima Atashin}, 
\ead{nima.atashin@eng.ui.ac.ir}
\author[CorAuth]{Behrouz Tork Ladani},
\ead{ladani@eng.ui.ac.ir} and %
\author{Mohammadreza Sharbaf}
\ead{m.sharbaf@eng.ui.ac.ir}

\address[a1]{Faculty of Computer Engineering, University of Isfahan, Isfahan, Iran}

\corauth[CorAuth]{Corresponding author.} 

\begin{abstract}

Detecting security vulnerabilities in open-source software is a critical task that is highly regarded in the related research communities. Several approaches have been proposed in the literature for detecting vulnerable codes and identifying the classes of vulnerabilities. However, there is still room to work in explaining the root causes of detected vulnerabilities through locating vulnerable statements and the discovery of paths leading to the activation of the vulnerability. While frameworks like SliceLocator offer explanations by identifying vulnerable paths, they rely on rule-based sink identification that limits their generalization. In this paper, we introduce VulPathFinder, an explainable vulnerability path discovery framework that enhances SliceLocator's methodology by utilizing a novel Graph Neural Network (GNN) model for detecting sink statements, rather than relying on predefined rules. The proposed GNN captures semantic and syntactic dependencies to find potential sink points (PSPs), which are candidate statements where vulnerable paths end. After detecting PSPs, program slicing can be used to extract potentially vulnerable paths, which are then ranked by feeding them back into the target graph-based detector. Ultimately, the most probable path is returned, explaining the root cause of the detected vulnerability. We demonstrated the effectiveness of the proposed approach by performing evaluations on a benchmark of the buffer overflow CWEs from the SARD dataset, providing explanations for the corresponding detected vulnerabilities. The results show that VulPathFinder outperforms both original SliceLocator and GNNExplainer (as a general GNN explainability tool) in discovery of vulnerability paths to identified PSPs. 

\end{abstract}


\makeatother

\end{frontmatter}

\addtolength{\parskip}{2mm}


\section{Introduction} \label{sec:intro}

Modern software systems are increasingly exposed to security vulnerabilities. Many of these are reported through the Common Vulnerabilities and Exposures (CVE) database~\cite{nvd2020}.  To defend against these threats, researchers have developed different automated vulnerability detection methods. Graph-based methods, in particular, have shown superior success due to their ability to capture the structural and semantic dependencies in code~\cite{chakraborty2021deep}. Despite their effectiveness in detecting vulnerable code, most current graph-based models act as black boxes, offering little to no insight into why a particular code is flagged as vulnerable. Without such an explanation, it would be difficult for developers to debug and mitigate detected flaws.

Vulnerability detection techniques can generally be grouped into two main categories: rule-based methods, which include both static and dynamic analysis, and data-driven approaches~\cite{harzevili2023survey}. Because it is difficult to define vulnerabilities, rule-based methods suffer from high false-positive rates, especially on complex code~\cite{harzevili2023survey}. In contrast, data-driven methods such as deep learning have emerged as powerful alternatives capable of generalizing from large code corpora. This capability is enabled by the extensive availability of open-source vulnerability data, which provides a rich foundation for training and analysis~\cite{osv2020}. Data-driven approaches can learn the latent information from vulnerable patterns and have shown better performance compared to static tools that utilize predefined rules~\cite{harzevili2023survey}. 

Among data-driven approaches, both sequence-based and graph-based approaches have been widely explored~\cite{harzevili2023survey}. Sequence-based methods serialize code into tokens and apply neural networks to identify vulnerability patterns. Graph-based models have proven effective by representing code as abstract syntax trees (ASTs), control-flow graphs (CFGs), or program dependence graphs (PDGs), enabling them to capture structural and semantic code dependencies~\cite{yamaguchi2014modeling}. However, despite their success, these models often yield coarse-grained predictions and lack transparency, making it difficult for developers to understand why a function or code snippet is flagged as vulnerable. This black-box nature poses significant challenges for analyzing root cause, trust, and fixing. 

To address the limitation mentioned above, we propose VulPathFinder, a Graph Neural Network (GNN)-based approach for identifying most probable paths from the potential sources to the detected vulnerability sink statements. VulPathFinder enhances the vulnerability path discovery method used by SliceLocator~\cite{cheng2024slicelocator} by utilizing a GNN model to first detect potential sink points (PSPs), i.e., the statements that are more likely to be the last chain of a vulnerable trace in the code. Unlike rule-based methods such as SliceLocator, which consider a set of predefined rules to identify candidate sink points, our method will be context-aware and capable of generalizing to unseen sink statements. Indeed, by training a GNN model to find PSPs, VulPathFinder better captures complex vulnerability patterns, retaining control and data dependencies between statements that might not be covered by rule-based approaches. After finding PSPs, inspired by the SliceLocator, we perform backward slicing starting from each of the sink points in the list. As a result, we will have a list of candidate paths leading to a sink point that make some corresponding subgraphs. Subgraphs are then fed into off-the-shelf graph-based detectors to compute their likelihood of being vulnerable. The subgraph with the most likelihood of being vulnerable is finally chosen. It shows the corresponding best candidate vulnerable path to be considered as the explanation of the detected vulnerability.

To evaluate the performance of the proposed model for sink point detection, we used a set of standard classification metrics. Moreover, to show the end-to-end performance of the explanation method (explainability) against the rival methods, we used the Triggering Line Coverage (TLC) metric~\cite{cheng2024slicelocator} to compare the achieved results with the original SliceLocator as well as GNNExplainer~\cite{ying2019gnnexplainer}. The latter is a model-agnostic explanation method for GNNs that is most influential for a given prediction. The results achieved show that VulPathFinder not only brings in acceptable precision and recall in sink point detection but also brings in higher end-to-end performance in terms of TLC that shows better explainability.

The rest of the paper is organized as follows: \autoref{sec:related} reviews the related work in conventional static and dynamic approaches, deep learning, and explainable AI approaches. In \autoref{sec:Approach}, the proposed method is explained. In \autoref{sec:Experiment}, experimental setup, evaluation metrics, and implementation details are explained. The results are shown in \autoref{sec:Results}. Limitations are addressed in \autoref{sec:Limitations}, and finally, we conclude the paper in \autoref{sec:conclusion}.

\section{Related Work} \label{sec:related}

\subsection{Conventional Static and Dynamic approaches} \label{Static and Dynamic approaches}

Static analysis tools such as CodeQL~\cite{codeql2023} and FindBugs~\cite{hovemeyer2004finding} use fixed rules to find vulnerabilities without executing the code; however, they suffer from high false positives and may miss complex vulnerabilities because defining vulnerable patterns is a challenging task~\cite{harzevili2023survey}. Dynamic analysis tools such as Valgrind~\cite{nethercote2007valgrind} and AddressSanitizer~\cite{asan2023} find vulnerabilities at runtime, but they depend on test cases and may miss unexecuted paths.

\subsection{Deep learning-based approaches} \label{Deep learning-based approaches}

The use of deep learning for detecting vulnerable functions and code snippets has increased rapidly in recent years, thanks to the abundant vulnerable open-source datasets~\cite{osv2020}. Graph Neural Networks (GNNs), in particular, have shown strong capability in capturing patterns inside graphs and have been widely applied to tasks such as traffic analysis~\cite{jiang2022graph} and social network modeling~\cite{kipf2016semi}. By representing source code as a graph, graph-based models can be leveraged to find intrinsic semantic and structural patterns by retaining control and data dependency inside code~\cite{yamaguchi2014modeling}. There exist different graph representations, such as abstract syntax tree (AST), control flow graph (CFG), control dependence graph (CDG), data dependence graph (DDG), and code property graph (CPG). CPG integrates AST, CFG, CDG, and DDG to create a unified view that encodes the syntactic and semantic dependencies~\cite{yamaguchi2014modeling}. Some works have used solely the sequence of tokens as their code representation, but by mapping code to a graph $G = (V, E)$, where $V$ are nodes which denote entities like variables or statements, and $E$ are edges inside the graph which show dependencies between two entities, we can better represent dependencies among statements. 

Several works have used GNN, such as Devign~\cite{zhou2019devign}, which is a Gated Graph Recurrent Network-based method that represents source code in a composite graph of ASTs, CFGs, and DFGs. Reveal~\cite{chakraborty2021deep} first extracts rich syntax-semantic features using Gated Graph Neural Network and embeds these features via code property graphs, then maximizes the separation between vulnerable and non-vulnerable code in real-world datasets using representation learning. DeepWukong~\cite{cheng2021deepwukong} first generates the CFG and Variable Flow Graph (VFG) to construct a PDG. It then conducts forward and backward traversals to create an Extended Flow Graph (XFG). Then it converts statement tokens into Doc2Vec vectors. Subsequently, information from XFG edges, along with vectorized code tokens, is used as input for k-GNNs. IVDetect~\cite{li2021vulnerability} is an interpretable method that first produces sub-token sequences of the code and then leverages BGRU with 
an attention mechanism to integrate ASTs, variable names, type features, sub-token sequences, and data/control dependencies into a comprehensive code representation. The representation is then processed by the Feature-Attention Graph Convolutional Network (GCN) model for training. Then it employs an explanation model called GNNExplainer~\cite{ying2019gnnexplainer} to identify critical sub-graphs as the explanation.

\subsection{Explanation approaches} \label{Explanation approaches}

Despite the effectiveness of GNN-based detectors at flagging vulnerable code, the interpretations and explanations of the cause of vulnerability remain unknown; these models just output a prediction score for each input without explaining the cause of the prediction. Recently, explainable AI (XAI) has emerged to address this gap~\cite{gunning2019xai}. Several techniques, such as GNNExplainer~\cite{ying2019gnnexplainer}, CFExplainer~\cite{lucic2022cf}, demonstrate the cause of predictions yielded by models by highlighting parts of the input that influence model outputs. GNNExplainer learns a minimal subgraph and a subset of node features that alone are sufficient to yield the same prediction as the full graph~\cite{ying2019gnnexplainer}. CFExplainer is a counterfactual explanation that identifies the smallest modifications to a graph's structure needed to reverse the model's prediction~\cite{lucic2022cf}. In the scope of vulnerability detection, it highlights which structural modifications could transform a code snippet from being classified as vulnerable to non-vulnerable, or vice versa. However, these methods often struggle with granularity and usability when applied to complex source code. This is because these models capture the difference between vulnerable code and non-vulnerable code without capturing the intrinsic behavior of vulnerabilities and their execution paths, and a slight change in input results in drastically different explanations. Also, most of the explainers deal with the models themselves, ignoring insights about the taint tracking and slicing. So, static analysis concepts such as taint propagation and slicing can be a promising complement to explainers.

\section{The Proposed Approach} \label{sec:Approach}
In this section, we present our framework, VulPathFinder, which enhances vulnerability path discovery by utilizing a GNN model to detect PSPs. Previous works (including SliceLocator)~\cite{li2021sysevr, cheng2024slicelocator} considered predefined rules—such as those related to library/API call, array usage, pointer usage, and arithmetic operations—to locate candidate sink points. In contrast, we employ a data-driven approach to locate candidate vulnerability sink points. This is the most important difference between our work and previous ones. By training a GNN model to find PSPs, our method better captures complex vulnerability patterns, retaining control and data dependencies between statements that might not be covered by rule-based approaches~\cite{yamaguchi2014modeling}. VulPathFinder offers several advantages over rule-based approaches. First, by training a specific model to identify PSPs, we can have a context-aware model that can capture vulnerable patterns. Second, VulPathFinder can be generalized to find unseen sink points across various vulnerability types.

The overall framework is depicted in Figure~\ref{fig:vulpathfinder_framework}, which consists of four main phases: 

\begin{enumerate}
	\item  Training GNN model for Sink Point Detection,
	\item Identification of PSPs: In this phase, taking advantage of the trained GNN model from the previous step, we find a list of PSPs,
	\item  Flow Path Generation: By having the list of PSPs, we perform backward slicing starting from each of the sink points in the list. At the end of this phase, we will have a list of candidate paths leading to a sink point.,
	\item  Flow Path Selection: In the last phase, the prediction score of each path is separately fed into the graph-based detector, and the path with the prediction score closest to the prediction score of the whole original graph will be chosen as the vulnerable path and considered as the explanation of the vulnerability.
\end{enumerate}

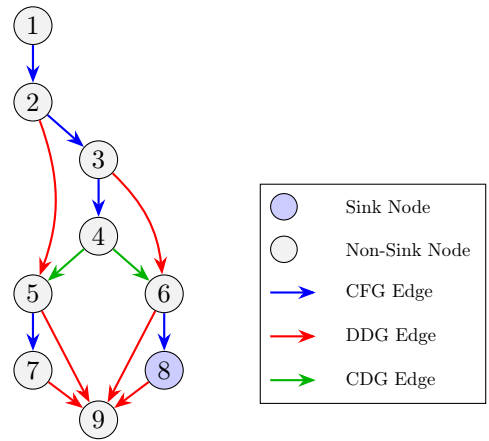
\begin{figure}[h]
    \centering
    \begin{tikzpicture}[
        node distance=5mm and 5mm,
        sink/.style={circle, draw=black, fill=blue!20, minimum size=5mm, inner sep=0pt},
        nonsink/.style={circle, draw=black, fill=gray!10, minimum size=5mm, inner sep=0pt},
        cfg/.style={-Stealth, blue, thick},
        ddg/.style={-Stealth, red, thick},
        cdg/.style={-Stealth, green!70!black, thick}
    ]
    \node[nonsink] (n1) {1};
    \node[nonsink, below=5mm of n1] (n2) {2};
    \node[nonsink, below right=4mm and 5mm of n2] (n3) {3};
    \node[nonsink, below=5mm of n3] (n4) {4};
    \node[nonsink, below left=4mm and 5mm of n4] (n5) {5};
    \node[nonsink, below right=4mm and 5mm of n4] (n6) {6};
    \node[nonsink, below=5mm of n5] (n7) {7};
    \node[sink, below=5mm of n6] (n8) {8};
    \node[nonsink, below=4mm of $(n7)!0.5!(n8)$] (n9) {9};

    \draw[cfg] (n1) -- (n2);
    \draw[cfg] (n2) -- (n3);
    \draw[cfg] (n3) -- (n4);
    \draw[cfg] (n5) -- (n7);
    \draw[cfg] (n6) -- (n8);

    \draw[ddg] (n2) to[bend left=20] (n5);
    \draw[ddg] (n3) to[bend left=20] (n6);
    \draw[ddg] (n5) -- (n9);
    \draw[ddg] (n6) -- (n9);
    \draw[ddg] (n7) -- (n9);
    \draw[ddg] (n8) -- (n9);

    \draw[cdg] (n4) -- (n5);
    \draw[cdg] (n4) -- (n6);

    \matrix [draw, right=10mm of n6, column sep=3mm, row sep=2mm, nodes={anchor=west, scale=0.7}] {
      \node[sink] {}; & \node {Sink Node}; \\
      \node[nonsink] {}; & \node {Non-Sink Node}; \\
      \draw[cfg] (0,0) -- +(.6,0); & \node {CFG Edge}; \\
      \draw[ddg] (0,0) -- +(.6,0); & \node {DDG Edge}; \\
      \draw[cdg] (0,0) -- +(.6,0); & \node {CDG Edge}; \\
    };
    \end{tikzpicture}
    \caption{Control flow graph with CFG, DDG, and CDG edges.}
    \label{fig:control_flow}
\end{figure}

Figure~\ref{fig:buffer_overflow} shows an example of a buffer overflow function, and its corresponding CPG is illustrated in Figure~\ref {fig:control_flow}.  We can see that the sink line is line 8, and several paths can be extracted by performing backward slicing, such as 1 --> 5 --> 8, 1 --> 6 --> 8, 7 --> 8, etc. So by ranking these paths based on their prediction score, we can select the path with the highest score as the explanation for the given vulnerable function.
Each step is detailed below:

\begin{figure*}[!t]
\centering
\begin{tikzpicture}[
    process/.style={rectangle, draw, fill=orange!10, text width=2.5cm, minimum height=0.7cm, align=center, font=\scriptsize},
    data/.style={rectangle, draw, dashed, fill=green!5, text width=2.0cm, minimum height=0.6cm, align=center, font=\scriptsize},
    input/.style={rectangle, draw, fill=yellow!10, text width=2.0cm, minimum height=0.6cm, align=center, font=\scriptsize},
    output/.style={rectangle, draw, fill=red!10, text width=2.0cm, minimum height=0.6cm, align=center, font=\scriptsize, very thick},
    detector/.style={rectangle, draw, fill=blue!10, text width=2.0cm, minimum height=0.6cm, align=center, font=\scriptsize},
    line/.style={-Stealth, thick},
    node distance=0.4cm and 1cm,
]

\node[data] (sard_dataset) {SARD Dataset};
\node[process, below=0.5cm of sard_dataset] (gnn_training) {GNN Training};
\node[data, below=0.5cm of gnn_training] (trained_gnn) {Trained GNN};

\node[input, right=2.5cm of sard_dataset] (input_code) {Input Code (C/C++)};
\node[process, below=0.5cm of input_code] (code_graph_gen) {Joern};
\node[data, below=0.5cm of code_graph_gen] (code_prop_graph) {Code Property Graph};
\node[process, below=0.5cm of code_prop_graph] (psp_ident) {Sink Points Detection};
\node[data, below=0.5cm of psp_ident] (list_psps) {List of PSPs};
\node[process, below=0.5cm of list_psps] (backward_slicing) {Backward Slicing};
\node[data, below=0.5cm of backward_slicing] (candidate_paths) {Candidate Paths};
\node[process, below=0.5cm of candidate_paths] (path_scoring) {Path Scoring};
\node[output, below=0.5cm of path_scoring] (vulnerable_path) {Most Probable Path};

\node[detector, right=2.5cm of path_scoring] (vuln_detector) {GNN Detector (e.g. Devign, Reveal, IVDetect)};

\draw[line] (sard_dataset) -- (gnn_training) -- (trained_gnn);
\draw[line] (input_code) -- (code_graph_gen) -- (code_prop_graph) -- (psp_ident)
            -- (list_psps) -- (backward_slicing) -- (candidate_paths) -- (path_scoring)
            -- (vulnerable_path);

\draw[line] (trained_gnn.east) -- ++(0.5,0) |- (psp_ident.west);
\draw[line] (vuln_detector.west) -- (path_scoring.east);

\begin{pgfonlayer}{background}
\node[draw, rounded corners, fill=gray!10, inner sep=0.3cm, 
fit=(sard_dataset) (gnn_training) (trained_gnn) 
(input_code) (code_graph_gen) (code_prop_graph) (psp_ident) (list_psps) 
(backward_slicing) (candidate_paths) (path_scoring) (vulnerable_path) (vuln_detector),
] {};
\end{pgfonlayer}

\end{tikzpicture}
\caption{Overview of the VulPathFinder Vulnerability Path Discovery Framework.}
\label{fig:vulpathfinder_framework}
\end{figure*}
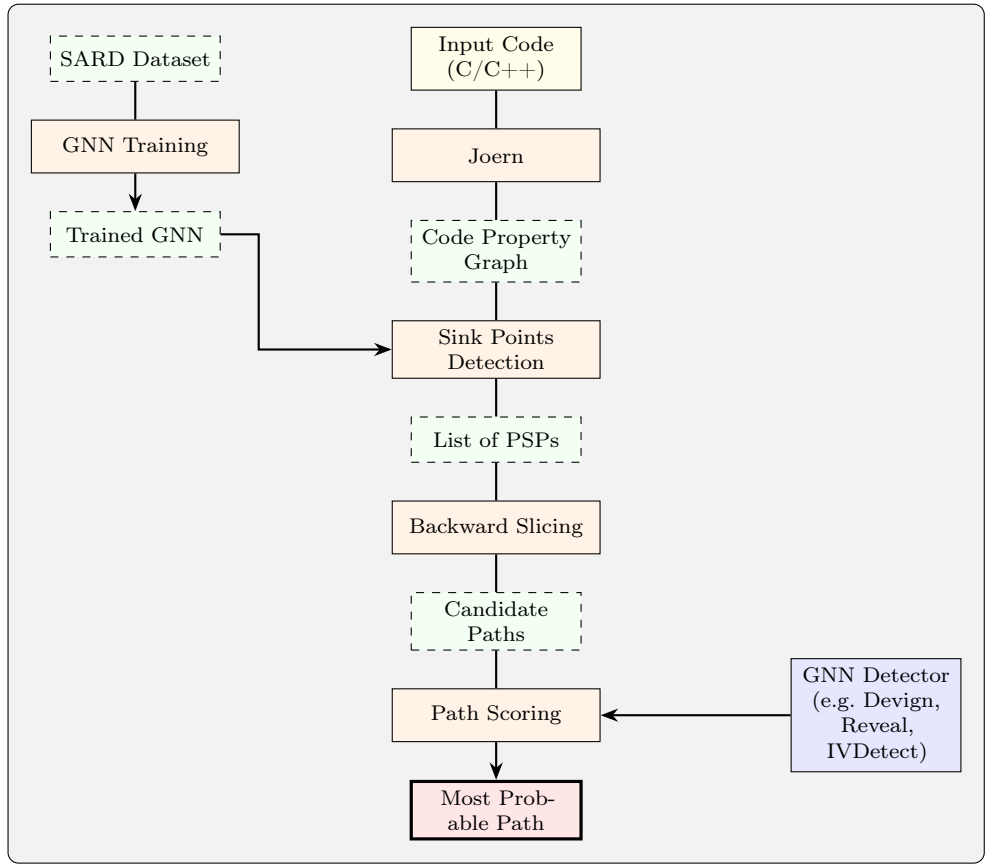

\definecolor{codegreen}{rgb}{0,0.6,0}
\definecolor{codegray}{rgb}{0.5,0.5,0.5}
\definecolor{codepurple}{rgb}{0.58,0,0.82}
\definecolor{backcolour}{rgb}{0.95,0.95,0.92}
\definecolor{vulncolor}{rgb}{0.8,0.1,0.1}

\lstdefinestyle{mystyle}{
    backgroundcolor=\color{backcolour},   
    commentstyle=\color{codegreen},
    keywordstyle=\color{blue},
    numberstyle=\tiny\color{codegray}, 
    stringstyle=\color{codepurple},
    basicstyle=\ttfamily\tiny, 
    breakatwhitespace=false,         
    breaklines=true,                 
    captionpos=b,                    
    keepspaces=true,                 
    numbers=left,                    
    numbersep=5pt,                  
    showspaces=false,                
    showstringspaces=false,
    showtabs=false,                  
    tabsize=2,
    frame=single,
    frameround=tttt,
    rulecolor=\color{black!30},
    xleftmargin=1.5em, 
    framexleftmargin=1em, 
    escapeinside={(*@}{@*)},
    moredelim=[is][\color{vulncolor}\bfseries]{❗}{❗}
}

\lstset{style=mystyle}

\begin{figure}[h]
    \centering
    \begin{lstlisting}[language=C, numbers=none]
void CWE121_Stack_Based_Buffer_Overflow()
{
(*@\raisebox{0pt}[0pt][0pt]{\llap{\normalfont\tiny\color{codegray}1\hspace{5pt}}}@*)    int * data;
(*@\raisebox{0pt}[0pt][0pt]{\llap{\normalfont\tiny\color{codegray}2\hspace{5pt}}}@*)    int * dataBadBuffer = (int *)ALLOCA(50*sizeof(int));
(*@\raisebox{0pt}[0pt][0pt]{\llap{\normalfont\tiny\color{codegray}3\hspace{5pt}}}@*)    int * dataGoodBuffer = (int *)ALLOCA(100*sizeof(int));
(*@\raisebox{0pt}[0pt][0pt]{\llap{\normalfont\tiny\color{codegray}4\hspace{5pt}}}@*)    if(globalReturnsTrueOrFalse())
    {
        
(*@\raisebox{0pt}[0pt][0pt]{\llap{\normalfont\tiny\color{codegray}5\hspace{5pt}}}@*)        data = dataBadBuffer;
    }
    else
    {
(*@\raisebox{0pt}[0pt][0pt]{\llap{\normalfont\tiny\color{codegray}6\hspace{5pt}}}@*)        data = dataGoodBuffer;
    }
    {
(*@\raisebox{0pt}[0pt][0pt]{\llap{\normalfont\tiny\color{codegray}7\hspace{5pt}}}@*)        int source[100] = {0};
(*@\raisebox{0pt}[0pt][0pt]{\llap{\normalfont\tiny\color{codegray}8\hspace{5pt}}}@*)        memmove(data, source, 100*sizeof(int));(*@\tikz[baseline=(char.base)]\node[rectangle,fill=vulncolor,text=white,font=\bfseries\tiny,inner sep=1pt] (char) {VULN};@*)
(*@\raisebox{0pt}[0pt][0pt]{\llap{\normalfont\tiny\color{codegray}9\hspace{5pt}}}@*)        printIntLine(data[0]);
    }
}
    \end{lstlisting}
    \caption{Buffer overflow example in C}
    \label{fig:buffer_overflow}
\end{figure}

\subsection{GNN Training for Sink Point Detection} \label{sec:GNN Training for Sink Point Detection}

\begin{figure}[!t]
\centering
\begin{tikzpicture}[
    layer/.style={rectangle, draw, rounded corners, minimum height=0.5cm, minimum width=2.8cm, align=center, fill=blue!10, font=\tiny},
    op/.style={rectangle, draw, minimum height=0.4cm, minimum width=2.8cm, align=center, fill=green!10, font=\tiny},
    arrow/.style={-Stealth, thick},
    input/.style={rectangle, draw, minimum height=0.5cm, minimum width=2.8cm, align=center, fill=yellow!10, font=\tiny},
    output/.style={rectangle, draw, minimum height=0.5cm, minimum width=2.8cm, align=center, fill=red!10, font=\tiny},
    skip/.style={-Stealth, thick, blue!50, dashed, bend right=45}
]


\node[input] (input) {Input (Features: 128, Edges)};

\node[layer, below=0.3cm of input] (gcn1) {GCNConv $(128 \to 256)$};
\node[op, below=0.2cm of gcn1] (bn1) {BatchNorm $(256)$};
\node[op, below=0.2cm of bn1] (relu1) {ReLU};
\node[op, below=0.2cm of relu1] (drop1) {Dropout $(0.5)$};

\node[layer, below=0.3cm of drop1] (gcn2) {GCNConv $(256 \to 256)$};
\node[op, below=0.2cm of gcn2] (bn2) {BatchNorm $(256)$};
\node[op, below=0.2cm of bn2] (relu2) {ReLU};
\node[op, below=0.2cm of relu2] (drop2) {Dropout $(0.5)$};

\node[below=0.2cm of drop2, font=\large] (dots) {$\vdots$};

\node[layer, below=0.2cm of dots] (gcnN1) {GCNConv $(256 \to 256)$};
\node[op, below=0.2cm of gcnN1] (bnN1) {BatchNorm $(256)$};
\node[op, below=0.2cm of bnN1] (reluN1) {ReLU};
\node[op, below=0.2cm of reluN1] (dropN1) {Dropout $(0.5)$};

\node[layer, below=0.3cm of dropN1] (gcnN) {GCNConv $(256 \to 2)$};

\node[output, below=0.3cm of gcnN] (output) {Output (2 classes)};

\draw[arrow] (input) -- (gcn1);
\draw[arrow] (gcn1) -- (bn1);
\draw[arrow] (bn1) -- (relu1);
\draw[arrow] (relu1) -- (drop1);
\draw[arrow] (drop1) -- (gcn2);
\draw[arrow] (gcn2) -- (bn2);
\draw[arrow] (bn2) -- (relu2);
\draw[arrow] (relu2) -- (drop2);
\draw[arrow] (drop2) -- (dots);
\draw[arrow] (dots) -- (gcnN1);
\draw[arrow] (gcnN1) -- (bnN1);
\draw[arrow] (bnN1) -- (reluN1);
\draw[arrow] (reluN1) -- (dropN1);
\draw[arrow] (dropN1) -- (gcnN);
\draw[arrow] (gcnN) -- (output);

\draw[skip] (gcn1.east) to[bend left=30] (gcn2.east);
\draw[skip] (gcn2.east) to[bend left=30] (gcnN1.east);
\draw[skip] (gcnN1.east) to[bend left=30] (gcnN.east);

\begin{scope}[on background layer]
\node[
  fit=(gcn2)(drop2)(gcnN1)(dropN1)(dots),
  draw, dashed, inner sep=0.15cm, fill=gray!5,
  label={[font=\tiny]right:{Repeated x 4}}
] {};
\end{scope}

\node[below=0.2cm of output, font=\tiny, text width=2.5cm, align=center] (legend) {
\textcolor{blue!50}{---} Skip connections
};

\end{tikzpicture}
\caption{Illustration of the GNN (6-layer GCN) architecture with skip connections and repeated blocks.}
\label{fig:gnn-architecture}
\end{figure}
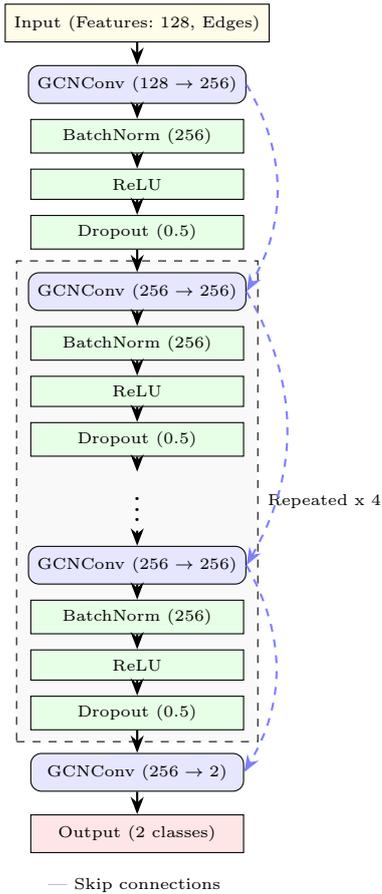

In order to train the GNN model, we used the Software Assurance Reference Dataset (SARD)~\cite{sard2020}. SARD provides ground-truth annotations for vulnerability-triggering statements; these were used to label corresponding CPG nodes as 'sink' (triggering points) or 'non-sink'. These labels are used in the training process as node labels for the node classification task to classify each node as either sink or non-sink. To achieve robustness and overcome class imbalance, we preprocessed the dataset to balance positive and negative samples to overcome the imbalance issue. We used a Graph Convolutional Network architecture, which uses message passing to capture dependencies among neighboring nodes in the CPG. We used 6 GCN layers, each followed by batch normalization and ReLU activation function to stabilize training and also to introduce nonlinearity. We also added Dropout with a probability of 0.5 after each hidden layer to prevent overfitting. As for node features, we trained 128-dimensional Word2Vec embeddings by using a random walk to encode node types (e.g., Identifier, CallExpression) and their content (e.g., variable name or function calls). The final output of the final layer is 1 or 0, representing sink or non-sink class for each node. Figure~\ref{fig:gnn-architecture} presents GNN's architecture.

Once the GNN model was trained and deployed to detect sink points, the approach proceeded as follows:

\subsection{Identification of PSPs} \label{sec:Potential Sink Points Identification}
We utilized the pretrained GNN model from the previous step to detect candidate sink points. At the end of this step, a list of PSPs is returned.

\subsection{Backward Slicing} \label{sec:Backward Slicing}
Inspired by SliceLocator~\cite{cheng2024slicelocator}, we generate a list of potential vulnerable paths by performing backward slicing from each of the predicted sink points in the previous step, all the way up to the source of the path. Then, we will have a list of candidate paths to further examine in the next step.

\subsection{Flow Path Scoring and Selection} \label{sec:Flow Path Generation}
Following the methodology of SliceLocator, the selection of the most probable path is determined by leveraging a target graph-based vulnerability detector such as Devign, Reveal, or IVDetect to assign an importance score to each path. Finally, the path with the highest importance score will be returned as the explanation of the vulnerable input code. To be precise, we calculate the probability of the given code graph $G$ as follows
\[ p_G = \Phi(\text{vec}(G)) \]
\noindent
where $\Phi$ represents the target detector model, and $\text{vec}$ denotes the Word2Vec embedding function that transforms $G$ into its vector representation. Then, for each path, we calculate the same probability, but this time only for the subgraph corresponding to that path. Then we calculate the importance score for each path as follows:
\[ \text{IS}_g = 1 - (p_G - p_g) \]
The closer the probability $p_g$  of each subgraph $g$ is to that of the original graph $G$, the higher the likelihood that the subgraph contains vulnerable statements.

\section{Experimental Evaluation} \label{sec:Experiment}

In this section, the experimental setup used to evaluate our approach is shown. We describe the dataset, the configuration of the training process, the metrics used for evaluation, and the baselines. All of the developed codes of VulPathFinder and datasets used in this work are available in our GitHub repository~\cite{vulpathfinder2025}.

\subsection{Dataset} \label{sec:Dataset}
For our experiment, we used the SARD dataset~\cite{sard2020}. We included six C/C++ weaknesses: CWE-121 to CWE-126, which are different sorts of buffer overflow. This selection resulted in a total of 9660 vulnerable functions, with each function containing multiple statements that are represented as nodes in our program graphs. Source code is parsed into graphs using Joern~\cite{joern2023} and SVF~\cite{svf2023}, with duplicates removed via MD5 hashing.

\subsection{Evaluation Metric} \label{sec:Evaluation Metric}
To evaluate the performance of our GNN model for sink point detection, we use standard classification metrics, including Precision, Recall, and F1-Score~\cite{harzevili2023survey}. To evaluate the end-to-end performance of our explanation method, we adopt the Triggering Line Coverage (TLC) metric, which is also used by the baseline method, SliceLocator, allowing for a fair comparison~\cite{cheng2024slicelocator}. TLC measures the overlap between the reported path, which serves as an explanation, and the actual ground truth statements that trigger the vulnerability.
TLC is calculated with the following equation:
\[ \text{TLC} = \frac{|s^{\text{e}} \cap s^{\text{v}}|}{|s^{\text{v}}|} \]

\noindent
where $s^e$ denotes the set of statements in the predicted vulnerable path and $s^v$ represents the set of labeled triggering statements as ground truths.

\subsection{Target Vulnerability Detectors} \label{sec:Target Vulnerability Detectors}
To thoroughly evaluate VulPathFinder's ability to provide explanations for different black-box models, we adopted three state-of-the-art graph-based vulnerability detectors as our targets: Devign~\cite{zhou2019devign}, Reveal~\cite{chakraborty2021deep}, and IVDetect~\cite{li2021vulnerability}. These models were chosen because they represent prominent deep learning approaches for vulnerability detection and were also utilized as target detectors in the SliceLocator study, allowing for direct comparison~\cite{cheng2024slicelocator}. For each of these detectors, we used their publicly available implementation. These models then served as the 'black-box' detectors for which VulPathFinder generated explanations for the vulnerability path discovery task.

\subsection{Baselines} \label{sec:Baselines}
We compare VulPathFinder against two baselines to benchmark its performance in providing vulnerability explanations:

SliceLocator: A state-of-the-art technique that employs a rule-based approach to identify PSPs and then uses backward slicing to generate explanations for vulnerabilities~\cite{cheng2024slicelocator}.

GNNExplainer: A model-agnostic explanation method for GNNs that identifies a critical subgraph that is most influential for a given prediction~\cite{ying2019gnnexplainer}.

\subsection{Implementation and Training Details} \label{sec:Implementation and Training Details}
The dataset was partitioned into training (70\%), validation (10\%), and test (20\%) sets. Class imbalance was addressed through a combination of oversampling the minority class in the training data and using a weighted loss function during training. All models were trained on a single NVIDIA RTX 3070ti GPU with a batch size of 64, using the Adam optimizer. 

\begin{figure}[!t]
    \raggedleft
    \includegraphics[width=1\columnwidth]{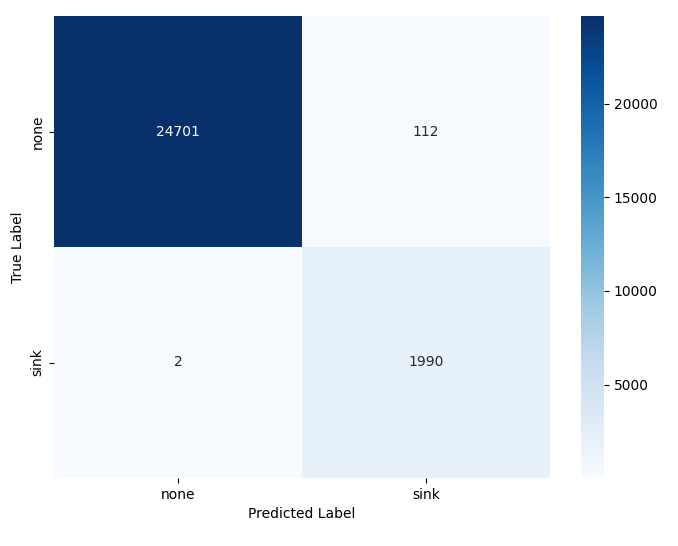}
    \caption{Confusion matrix on the test set.}
    \label{fig:confusion_matrix}
\end{figure}

\section{Results} \label{sec:Results}
In this section, we present the experimental results of our evaluation. First, we report the performance of our GNN model for sink point detection, followed by results for vulnerability path discovery, comparing VulPathFinder against the baselines.

\subsection{Sink Point Detection Performance} \label{sec:Sink Point Detection Performance}

We first evaluated our trained GNN model on the task of classifying graph nodes as sinks. The performance of the GNN model on the test set is summarized in Table \ref{tab:metrics}. The model achieved a high precision of 0.97 and a macro F1-score of 0.98. The model's ability to detect the majority of true sink nodes is highlighted by its 0.99 recall score. This high recall is required because the correct vulnerable path cannot be included for analysis if its sink is not identified. The confusion matrix is shown in Figure \ref {fig:confusion_matrix}, and it confirms this low rate of false negatives for the sink class.

\subsection{Vulnerability Explanation Performance} \label{sec:Vulnerability Localization Performance}

In the second part of our evaluation, we assessed the end-to-end performance of VulPathFinder in explaining vulnerabilities against the baselines. Table \ref{tab:approach_model_matrix} shows the average TLC scores across the test set for all methods. VulPathFinder achieved an average TLC score of 98\%, outperforming both SliceLocator and GNNExplainer. Although SliceLocator achieves a respectable average TLC of 92\%, its rule-based nature of sink identification prevents it from generalizing to unforeseen vulnerability patterns. GNNExplainer shows the lowest performance, with an average TLC of 81\%. The reason for the low performance of GNNExplainer can be attributed to the lack of explicit modeling of taint flow and dependencies within code, which are important for understanding many vulnerabilities. This result highlights a key challenge for applying general-purpose XAI techniques in the domain of software security. This observation validates the need to incorporate program analysis concepts, such as slicing, to give insightful explanations of software vulnerabilities.

\begin{table}[h]
  \centering
  \caption{Model Performance Metrics}
  \label{tab:metrics}
  \begin{tabular}{|l|c|}
    \hline
    \textbf{Metric} & \textbf{Value} \\ \hline
    Precision & 0.97 \\ \hline
    Recall    & 0.99 \\ \hline
    F1-Macro  & 0.98 \\ \hline
  \end{tabular}
\end{table}

\begin{table}[!t]
  \centering
  \caption{Comparison of TLC scores (explanation power) of different approaches with three underlying  state-of-the-art graph-based vulnerability detectors}
  \label{tab:approach_model_matrix}
  \begin{tabular}{|l|c|c|c|}
    \hline
    \textbf{Approach} & \textbf{IVDetect} & \textbf{Devign} & \textbf{Reveal} \\ \hline
    VulPathFinder     & 0.98               & 0.99            & 0.98            \\ \hline
    SliceLocator      & 0.90               & 0.97            & 0.91            \\ \hline
    GNNExplainer      & 0.71               & 0.86            & 0.86            \\ \hline
  \end{tabular}
\end{table}

\section{Limitations and Threat to Validity} \label{sec:Limitations}

First, the SARD dataset we used is an academic dataset that includes synthetic code that might not be used in real-world software programs~\cite{sard2020}. Second, we only evaluated 6 types of CWEs that are mostly related to buffer overflow vulnerability. Third, we only evaluated on c/c++ codes, although we can easily extend this work to use more programming languages such as Java, Python, etc.

\section{Conclusion}\label{sec:conclusion}
In this paper, we introduced VulPathFinder, a GNN-based framework for explainable vulnerability path discovery that outperforms traditional rule-based methods. By training a dedicated GNN model to identify potential sink points (PSPs) in code, our approach moves beyond the limitations of fixed heuristics and learns to recognize complex, context-aware vulnerability patterns. By integrating this learned sink detection with program slicing and path ranking, VulPathFinder successfully identifies and highlights the most probable vulnerable execution paths, providing developers with actionable insights.

Our experiments on the SARD dataset demonstrate the superiority of this data-driven approach. The sink detection model achieved high precision and recall, and the full VulPathFinder framework significantly outperformed both the rule-based SliceLocator and the general-purpose GNNExplainer in terms of Triggering Line Coverage. This work underscores the potential of combining deep learning with program analysis principles to build not only accurate but also interpretable tools for software vulnerability analysis.

\bibliographystyle{unsrtnat}
\bibliography{biblio}

\begin{wrapfigure}[8]{l}{1.5cm}
	\begin{center}
		\includegraphics[width=2.3cm, height= 2.85cm]{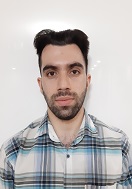}
	\end{center}
\end{wrapfigure}
\noindent
\newline
{\bf Nima Atashin} {received his bachelor’s degree in Computer Engineering from Isfahan University of Technology in 2022. He is currently pursuing an M.Sc. in Software Engineering in the Faculty of Computer Engineering at the University of Isfahan. His research interests include explainable AI, graph neural networks, and software vulnerability detection.}

\begin{wrapfigure}[8]{l}{1.5cm}
	\begin{center}
		\includegraphics[width=2.3cm, height= 2.85cm]{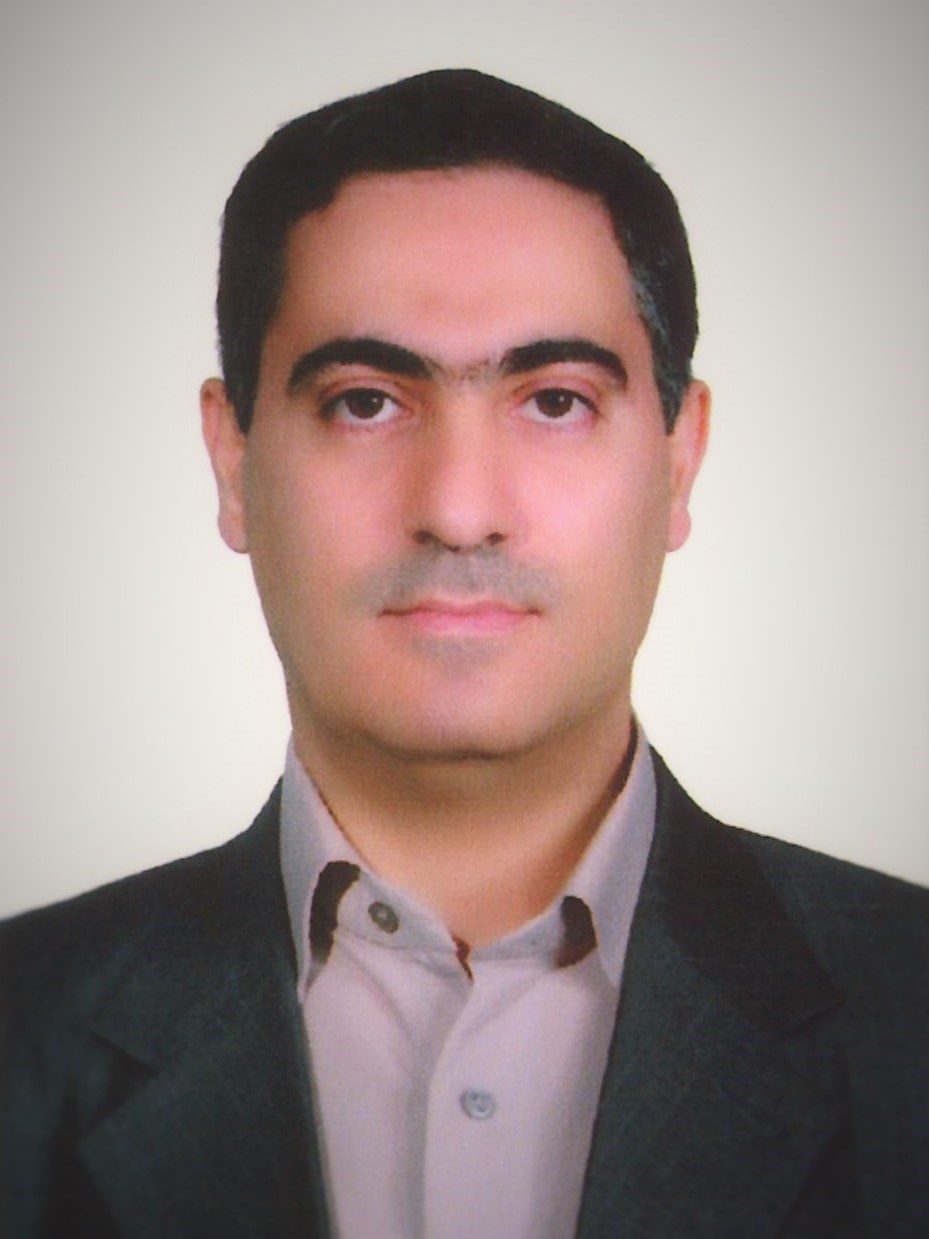}
	\end{center}
\end{wrapfigure}
\noindent
\newline
{\bf Behrouz Tork Ladani} {received his bachelor’s degree in computer engineering from the University of Isfahan (UI), Isfahan, Iran, in 1996, M.Sc. degree in software engineering from the Amirkabir University of Technology, Tehran, Iran, in 1998, and Ph.D. degree in software engineering from the University of Tarbiat Modarres, Tehran, Iran, in 2005. He joined UI in 2005, where he is currently a professor of Software Engineering. He is the author of more than 70 articles. His research interests are around modeling, analysis, and verification of security in information systems, including software security (vulnerability detection and malware analysis) and soft security (computational trust, rumor control, and opinion formation in social networks). Behrouz is a member of the Iranian Society of Cryptology (ISC), and he is the Managing Editor of the Journal of Computing and Security (JCS).}

\begin{wrapfigure}[8]{l}{1.5cm}
	\begin{center}
		\includegraphics[width=2.3cm, height= 2.85cm]{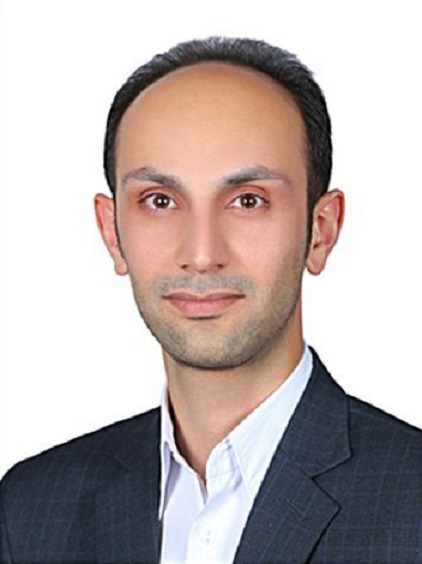}
	\end{center}
\end{wrapfigure}
\noindent
\newline
{\bf Mohammadreza Sharbaf} {is an Assistant Professor in Computer Engineering at the University of Isfahan (UI). He is interested in Model-Driven Software Engineering, Collaborative Modeling, Low-Code Development Platforms, Software Development Methodologies, Design Patterns, and Semantic Web (Semantic Reasoning). His current research is focused on software testing, inconsistency management, and multi-view modeling. Mohammadreza received his B.Sc. from the Isfahan University of Technology, Isfahan, Iran, in 2013, and his M.Sc. and Ph.D from the UI, Isfahan, Iran, in 2016 and 2022, both in Software Engineering. Now, he is the director of Model-Driven Software Engineering Research Group (MDSERG) at UI.}

\end{document}